\documentclass[aps,prd,preprint,groupedaddress,showpacs]{revtex4}
\usepackage{amssymb,amsmath,graphicx}

\usepackage{verbatim,color,ulem}

\begin{document}
\title{Brownian motion of a charged particle in electromagnetic fluctuations at finite temperature}
\author{Jen-Tsung Hsiang}
\email{cosmology@mail.ndhu.edu.tw}
\author{Tai-Hung Wu}
\author{Da-Shin Lee}
\email{dslee@mail.ndhu.edu.tw}
\affiliation{Department of Physics, National Dong Hwa University,
Hualien, Taiwan, R.O.C.}

\date{\today}

\begin{abstract}
The fluctuation-dissipation theorem is a central theorem in
nonequilibrium statistical mechanics by which the evolution of
velocity fluctuations of the Brownian particle under a fluctuating
environment is intimately related to its dissipative behavior. This
can be illuminated in particular by an example of Brownian motion in
an ohmic environment where the dissipative effect can be accounted
for by the first-order time derivative of the position. Here we
explore the dynamics of the Brownian particle coupled to a
supraohmic environment by considering the motion of a charged
particle interacting with the electromagnetic fluctuations at finite temperature.
We also derive particle's equation of motion, the Langevin equation, by minimizing the
corresponding stochastic effective action, which is obtained with
the method of Feynman-Vernon influence functional. The
fluctuation-dissipation theorem is established from first
principles. The backreaction on the charge is known in terms of
electromagnetic self-force given by a third-order time derivative of
the position, leading to the supraohmic dynamics. This self-force
can be argued  to be insignificant throughout the evolution when the
charge barely moves. The stochastic force arising from the
supraohmic environment is found to have both positive and negative
correlations, and it drives the charge into a fluctuating motion.
Although positive force correlations give rise to the growth of the
velocity dispersion initially, its growth slows down when
correlation turns negative, and finally halts, thus leading to the
saturation of the velocity dispersion. The saturation mechanism in a
suparohmic environment is found to be distinctly
 different from that in an
ohmic environment. The comparison is discussed.
\end{abstract}

 \maketitle

\allowdisplaybreaks

\section{Introduction}
One of the fundamental problems in statistical mechanics concerns
the microscopic origin of dissipation and relaxation of a
nonequilibrium system in its course toward equilibrium with the
thermal bath. It is known that the particle moving in a randomly
fluctuating medium undergoes Brownian motion. A phenomenological but
rather successful approach to study the evolution of Brownin motion
is by means of the Langevin equation. This Langevin equation is a
classical equation of motion by phenomenologically adding the terms
that account for the effects of dissipation and fluctuations under
the influence of a fluctuating environment. These two effects are
ultimately responsible for evolving  into thermodynamic equilibrium
of the Brownian particle with the bath, and are thus related by the
fluctuation-dissipation theorem (FDT). The FDT is one of the
cornerstones of statistical mechanics~\cite{PA}. The {\it static}
version of the FDT~\cite{Bh} relates the response of a system in
equilibrium to action of a small external perturbation. The result
of it gives a relation between noise correlations and susceptibility
in frequency domain and the proportionality constant depends on
temperature.  The extension of the {\it static} FDT to the situation
away from equilibrium can be realized in terms of the Langevin
equation.  The Einstein relation, one of the manifestations of the
{\it dynamic} FDT, links the effects between friction and noise
correlations in the Langevin equation, and thus plays an essential
role in stabilizing the dynamics of the Brownian particle. The most
striking feature of this relation is that it can determine the
irreversible evolution of the particle from the fluctuations
correlation of the heat bath. Thus, the origin of this
irreversibility of the Brownian dynamics can be attributed to the
fluctuating nature of the environment.

A very clear microscopic description,  leading to the Langevin
equation, within the context of  one-particle quantum mechanics has
been presented by Caldeira and Leggett.  They considered a specific
system-environment model that the particle interacts with an
environment composed of a infinite number of harmonic oscillators by
linear coupling of the oscillator and particle
coordinates~\cite{CAL}. If the quantum states of harmonic
oscillators are thermally distributed, the relative importance
between quantum and thermal fluctuations depends on the temperature
under consideration. The effects of environmental degrees of freedom
on the particle can be investigated with the method of
Feynman-Vernon influence functional by integrating out environment
variables within the context of the closed-time-path
formalism~\cite{CAL,SC,FE,GR,HPZ}. The more complicated interaction
by considering nonlinear couplings of the particle coordinate can
also be studied perturbatively in~\cite{HPZ2}. Under the classical
approximation where the intrinsic quantum fluctuations of the
particle is ignored, the  Langevin equation can be obtained by
minimizing the corresponding effective stochastic action.

At low temperatures, the particle is affected mainly by quantum
fluctuations of the environment, and it will lead to a new
phenomenon, the so-called quantum Brownian motion. Many
experimentally accessible systems (such as dissipation in quantum
tunneling)~\cite{CAL}, and the problems of quantum measurement
theory (such as quantum decoherence of the system due to the
interaction with its environment)~\cite{ZU} share the similar
dynamics as quantum Brownian motion.  At high temperatures  in which
the thermal fluctuations dominate, the problem of Brownian motion is
described by the classical dynamics introduced in the previous
paragraph. The relativistic Brownian motion has also been discussed
in~\cite{HU3}.

Here we would like to stress that under the approach of
Feynman-Vernon influence functional, the corresponding FDT can be
{\it derived} from  first principles  for a given microscopic model
in terms of the Green's functions of  environment variables. The FDT
is a central theorem in nonequilibrium statistical mechanics by
which the evolution of velocity fluctuations of the particle under a
fluctuating environment is intimately related to its dissipative
behavior. Thus, in this paper, we wish to explore the dynamics of
the Brownian particle in the supraohmic environment where
backreaction dissipation is governed by the term  with high-order
time derivatives of the position than the one (a first-order time
derivative) in a ohmic case. The known example is the motion of the
charged particle under the electromagnetic fluctuations at finite
temperature. The non-uniform motion of the charge will emit
radiation that backreacts on itself through the electromagnetic
self-force given by a third-order time derivative of the position.
The stochastic noise, which encodes the influence of quantum/thermal
statistics of the fields, drives the charge into a fluctuating
motion~\cite{JL,HWL}. It is also known that, in nonrelativistic motion, the dissipation term is
significant only on an extremely short time scale, which is the time for
light to travel across the classical radius of the charged particle, $\approx  10^{-23}\text{s} $ for an electron.  It is then of interest
to find out the mechanism by which the velocity
fluctuations saturate, when backreaction dissipation is negligible during the
evolution, that is, when the particle barely moves in a supraohmic
environment.

The dynamics of a charged particle coupled with the electromagnetic
fields has been studied quantum-mechanically in the
system-plus-environment approach. We treat the particle as the
system of interest, and the degrees of freedom of fields as the
environment. The effects of fields on the particle is then obtained
by integrating out field variables~\cite{SC,FE,CAL,GR,HPZ}. In this
approach, the decoherence phenomena of the charged particle under
the influence of electromagnetic vacuum fluctuations in the presence
of the conducting plate has been studied in~\cite{JL,HLW}. The
evolution of charge's velocity dispersion affected by quantum
electromagnetic fields near a conducting plate~\cite{HWL} and from
the electromagnetic squeezed vacuum~\cite{HWL1} are also studied. In
the latter works, we investigate the possibility of reducing the
velocity dispersion of the charged particle by tuning the parameters
in these quantum states. In~\cite{YF}, the authors have shown that
if the particle dose not  move significantly such that the
electromagnetic self-force can be ignored, then the velocity
dispersion {\it still} reaches a constant value at asymptotical
times. Thus, constrained by the uncertainty principle, it is quite
reasonable that the particle cannot extract energy indefinitely from
the vacuum state of environment fields. It results from the
fact that the integral of the correlation function of stochastic
forces, which will be defined later, over the whole time domain
vanishes. Similar results are also obtained in Ref.~\cite{FR}. As
will be seen later, it will also lead to the saturation of velocity
fluctuations of a Brwonian particle under a supraohmic environment
at {\it finite temperature}. This work is a follow-up from our earlier investigations\cite{HWL,HWL1}, where more details on derivation can be found.

The Lorentz-Heaviside units with $\hbar=c=1$ will be adopted unless otherwise noted. The metric is $\eta^{\mu\nu}=\text{diag}(+1,-1,-1,-1)$.

\section{Langevin equations}\label{sec1}
The dynamics of a nonrelativistic point particle of charge $e$ interacting with the quantized electromagnetic fields can be described by the Lagrangian,
\begin{equation*}
    L[\mathbf{q},\mathbf{A}_{\mathrm{T}}]=\frac{1}{2}m\dot{\mathbf{q}}^2-\frac{1}{2}
    \int\!d^3\mathbf{x}\,d^3\mathbf{y}\;\varrho(x;\mathbf{q})G(\mathbf{x},\mathbf{y})\varrho(y;\mathbf{q})
    +\int\!d^3\mathbf{x}\left[\frac{1}{2}(\partial_\mu\mathbf{A}_{\mathrm{T}})^2+\mathbf{j}\cdot\mathbf{A}_{\mathrm{T}}\right]\,,
\end{equation*}
in terms of the transverse components of the gauge potential $\mathbf{A}_{\mathrm{T}}$, and the position ${\bf q}$ of the point charge in the Coulomb gauge. The instantaneous Coulomb Green's
function $G(\mathbf{x},\mathbf{y})$ satisfies the Gauss's law. The charge and the current densities take the form
\begin{equation}
    \varrho (x; {\bf q}(t)) =e\, \delta^{(3)} ( {\bf x}-{\bf q}(t) ) \,,\qquad\qquad
{\bf j} (x;  {\bf q}(t)) = e\,  \dot{\bf q} (t) \, \delta^{(3)} ( {\bf
x}-{\bf q}(t) ) \,. \label{charge-current}
\end{equation}
The density matrix of the particle-field system ${\hat \rho}(t)$ evolves unitarily according to
\begin{equation}
    {\hat \rho} (t_f) = U(t_f, t_i) \, {\hat \rho} (t_i) \, U^{-1} (t_f,
t_i )
\end{equation}
with $U(t_f,t_i)$ the time evolution operator. We assume that the state of the particle-field at an initial time $t_i$ is facterizable as ${\hat \rho} (t_i) = {\hat \rho}_{e} (t_i) \otimes {\hat \rho}_{{\bf A}_{\rm T}} (t_i)$, and that the electromagnetic fields are initially in thermal equilibrium at temperature $ T=1/\beta $ so that its density operator takes the form
\begin{equation}
\hat{\rho}_{{\bf A}_{\rm T}} (t_i) =e^{-\beta  H_{{\bf A}_{\rm T}}}/\operatorname{Tr}\left\{e^{-\beta  H_{{\bf A}_{\rm T}}}\right\}\,, \label{initialcondphi}
\end{equation}
where $ H_{{\bf A}_{\rm T}} $ is the Hamiltonian of the free
fields.

After integrating out the degrees of freedom of the fields, the Langevin equation is obtained~\cite{HWL},
\begin{equation}
m \ddot{q}^i +e^2\left(\delta^{i l} \frac{d}{dt} - \dot{q}^l (t)
\nabla_i\right) \int_{-\infty}^{\infty}dt' \; G_{R}^{lj} \left[{\bf
q}(t),{\bf q}(t'); t-t'\right]\,\dot{q}^j (t') =f_s^i(t)\, ,
\label{nonlinearlangevin}
\end{equation}
where
\begin{equation}
f_s^i (t) =-\hbar\, e \, \left( \delta^{il} \frac{d}{d t} - \dot{q}^l (t)\nabla^i \right) \, \xi^l (t)
\end{equation}
with the noise-noise correlation functions,
\begin{equation}
\langle\xi^i(t)\rangle =0\,,\qquad\qquad\langle\xi^i (t)\xi^j(t')\rangle=\frac{1}{\hbar} G_{H}^{ij} \left[{\bf q}(t), {\bf q}(t');
t-t'\right]\, , \label{noisecorrel}
\end{equation}
and
\begin{eqnarray}
    \hbar\,G_{R}^{ij}(x-x')&=&i\,\theta(t-t')\,\bigl<\left[A_{\mathrm{T}}^i(x),A_{\mathrm{T}}^j(x')\right]\bigr>\,,\label{commutator}\\
    \hbar\,G_{H}^{ij}(x-x')&=&\frac{1}{2}\,\bigl<\left\{A_{\mathrm{T}}^i(x),A_{\mathrm{T}}^j(x')\right\}\bigr>\,,\label{anticommutator}
\end{eqnarray}
are the Green's function of the electromagnetic potentials at finite
temperature. It is seen that the influence of the electromagnetic
fields are expressed by an integral of the dissipation kernel
$G_{R}^{ij}$ over the past history of charge's motion, and by a
stochastic noise $\boldsymbol{\xi}$ that drives the charge into a
fluctuating motion. As it stands, Eq.~\eqref{nonlinearlangevin} is a
nonlinear Langevin equation with non-Markovian backreaction, and the
noise depends in a complicated way on the charge's trajectory
because the noise correlation function itself is a functional of the
trajectory.

The fluctuation and dissipation effects of the electromagnetic
fields on the motion of the charged particle are associated with the
kernels $G_{H}^{ij}$ and $G_{R}^{ij}$ respectively. They in turn are
linked by the fluctuation-dissipation relation where the Fourier
transform of the fluctuation kernel $G_{H}^{ij}$ is related to the
imaginary part of the retarded kernel $G_{R}^{ij}$ as
follows~\cite{HWL}
\begin{equation}
G_{H}^{ij}[{\bf q}(t),{\bf
q}(t');\omega]=\operatorname{Im}\left\{G_{R}^{ij}[{\bf q}(t),{\bf
q}(t');\omega]\right\}\coth\left[\frac{\beta \hbar
\omega}{2}\right]\,. \label{df-T}
\end{equation}
This relation is established from first principles. The explicit
expression of the kernels $G_{H}^{ij}$ and $G_{R}^{ij}$ will be
introduced when they are needed later.

\section{velocity fluctuations}\label{sec2}
The nonlinear, non-Markovian Langevin equations are far too complicated to proceed further without any approximation. The appropriate approximation for nonrelativistic motion is the dipole approximation, which amounts to considering the backreaction solely from the electric fields. The Langevin equation under the dipole approximation reduces to
\begin{equation}\label{xxx}
    m\ddot{q}^i(t)+e^2\int_0^tdt'\;\dot{g}^{ii}_R(t-t')\,\dot{q}^i(t')=f_s^i (t)\,,
\end{equation}
from Eq.~\eqref{nonlinearlangevin} and
\begin{equation}
    f_s^i(t)=-\hbar\,e\,\dot{\xi}^i(t)\,.
\end{equation}
The retarded Green's function in the dipole approximation is denoted by $g_R$, and can be
expressed in terms of the spectral density $\rho$ as
\begin{equation}\label{see}
    g^{ij}_R(\tau)=-\theta(\tau)\int_0^{\infty}\frac{dk}{\pi}\;\rho^{ij}(k)\sin(k\tau)\,.
\end{equation}
In the isotropic thermal bath, the spectral density takes a simple
form~\cite{HWL}
\begin{equation}
    \rho^{ij}(k)=-\frac{k}{3\pi}\delta^{ij}\,. \label{speden}
\end{equation}
The accompanying noise-noise correlation functions due to the
fluctuations of the electric fields at finite temperature are given
by
\begin{equation}
    \langle f_s^i(t)\rangle=0\,,\qquad\qquad\langle f_s^i(t) f_s^j(t')\rangle=\hbar\,e^2\frac{\partial^2}{\partial t\partial t'}\,g^{ij}_{H}(t-t')\,,
\end{equation}
with
\begin{equation}\label{hadama-x}
    g^{ij}_{H}(\tau)=-\int_0^{\infty}\frac{dk}{2\pi}\;k^2\rho^{ij}(k)\coth\left[\frac{\beta\hbar k}{2}\right]\cos(k\tau)\,.
\end{equation}
Again the noise kernel $g^{ij}_{H}$ can be seen related to the dissipation kernel $g^{ij}_{R}$ via a fluctuation-dissipation relation under the dipole approximation, derived from Eq.~\eqref{df-T}. After carrying out the integration in Eq.~\eqref{xxx} the Langevin equation becomes physically more transparent,
\begin{equation}\label{ljsdfls}
     m_{r}\ddot{q}^i(t)-\frac{e^2}{6\pi}\dddot{q}^{i}(t)=f_s^i(t) \,
     .
\end{equation}
The non-uniform motion of the charge results in radiation that
backreacts on the charge itself through the electromagnetic
self-force. This backreaction occurs at the moment when radiation is
emitted~\cite{Roh,Roh2,HWL}. Thus, it may lead to short-distance divergence
in the coincidence limit due to the assumption of a point-like
particle. This ultraviolet divergence must be regularized to have a
finite and unambiguous result. The divergent part is absorbed by
particle mass renormalization, $m_{r}=m+e^2\Lambda/3\pi^2$, where $\Lambda$
is the energy cutoff scale related to the inverse of the charge's
wavepacket width. It essentially quantifies the intrinsic
uncertainty of the charged particle. Then the finite backreaction
effect is given by the well-known result, a third-order time
derivative of the position~\cite{JA}.

For a nonrelativistic particle, the introduced cutoff wavelength
should be much larger than its Compton wavelength, namely
$\Lambda^{-1}\gg\lambda_C=\hbar/m_rc^2$, and in turn much greater
than the classical radius of the charged particle, $r_e=e^2/m_rc$.
Thus, when the time scale associated with the cutoff frequency,
$t\sim2\pi/\Lambda$, is much longer than the characteristic time
scale $\tau_e\sim r_e/c$, an extremely short time scale,
the electromagnetic self-force can be safely ignored. If we assume
that the charged particle starts off from the rest at $t=0$, then
the solution to the Langevin equation \eqref{ljsdfls} has a very
simple form
\begin{equation}
    v_{i}(t)=\frac{1}{m_r}\int^{t}_{0}ds\;f_{s}^{i}(s)\,,
\end{equation}
and the corresponding velocity dispersion is given by
\begin{equation}
    \langle\Delta v_i^2(t)\rangle=\frac{1}{m_r^2}\int_0^tds\int_0^tds'\;\langle f_{s}^{i}(s)f_{s}^{i}(s')\rangle \,.
\end{equation}
We see that the evolution of velocity fluctuations is governed by the force-force correction function, whose finite temperature contribution $\langle f_s^i(\tau)f_s^i (0)\rangle_{\beta} $ is given by
\begin{align}
    C_{FF}(\tau)&=\langle f_s^i(\tau)f_s^i (0)\rangle_{\beta} \notag\\
                &=\frac{2}{\pi^2}\frac{\hbar\,e^2}{\beta^{4}\hbar^4}\operatorname{Re}\left\{\zeta(4,1+i\frac{\tau}{\beta \hbar})\right\}\,,
\end{align}
where its vacuum part has been subtracted. $\zeta (n,z)$ is the
$n$th derivative of the zeta function.
\begin{figure}
    \centering
        \scalebox{0.6}{\includegraphics{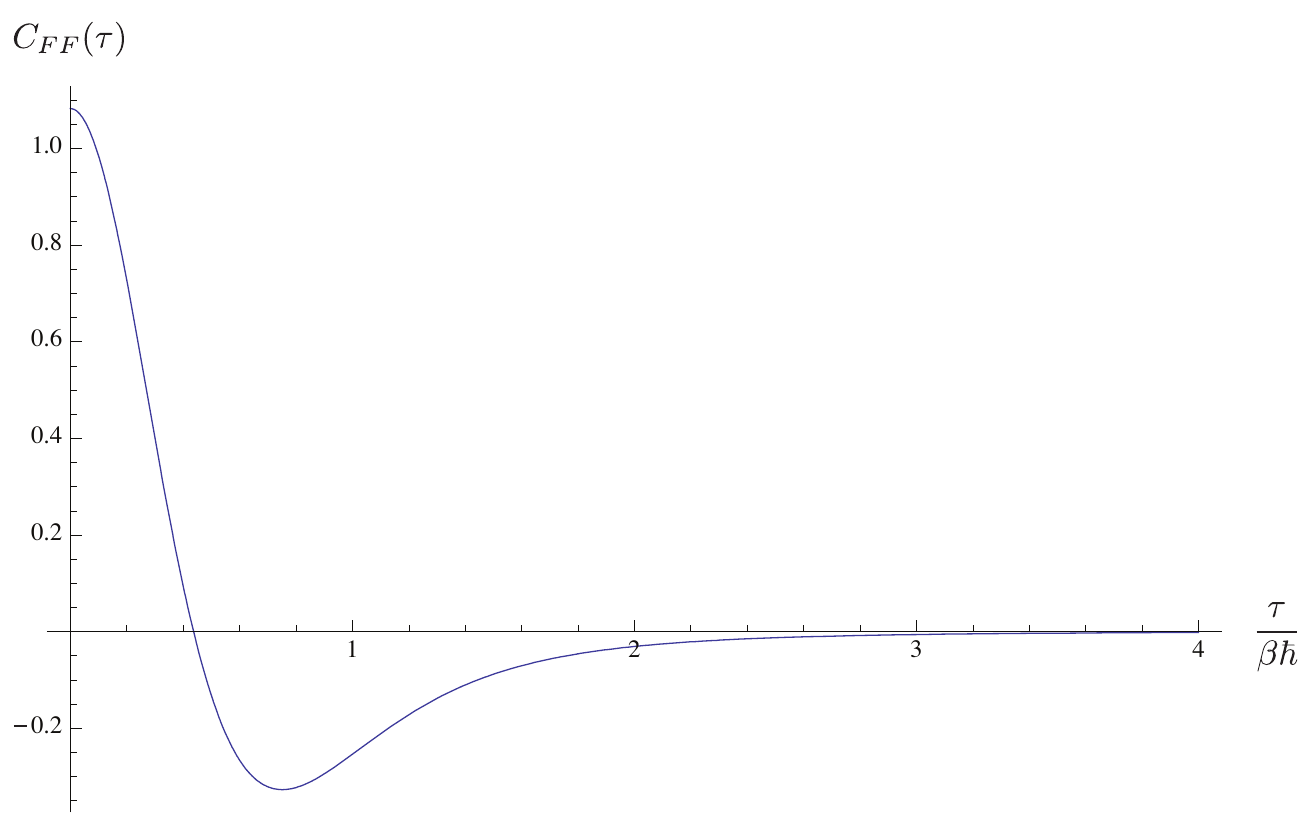}}
        \caption{The correlation function $C_{FF}(\tau)$ normalized with $\frac{2}{\pi^2}\frac{\hbar\,e^2}{\beta^{4}\hbar^4}$ is drawn as a function of  $\tau$  in units of $\frac{1}{\beta \hbar}$. }\label{Fi:Cffs}
\end{figure}
The correlation function $C_{FF}(\tau)$ is plotted against the time difference $\tau=t-t'$, scaled by $\beta\hbar$ in Fig.~\ref{Fi:Cffs}. It shows that the stochastic noise tends to have
positive correlation for small $\tau$, and then $C_{FF}(\tau)$ turns negative at $\tau\sim\mathcal{O}(\beta)$. It means that if we measure the electric field at certain time and find out the field points to one particular direction, then the moment $\tau$ later, within the time difference of order $\mathcal{O}(\beta)$, if we perform a similar measurement again at the same location, there is high probability that the direction of the electric field remains the same. However, when the time difference between measurements is greater than $\mathcal{O}(\beta)$, we probably have the result that the electric field points to the opposite direction instead of the same direction. Another way of understanding its significance is to compute the mean power
$\overline{P}_s(t)$ done by the stochastic force $f_s(t)$,
\begin{align}
    \overline{P}_s(t)=\langle f_s(t)v(t)\rangle_{\beta}&=\frac{1}{m_r}\int_0^tds\;\langle f_s(t)f_s(s)\rangle_{\beta}=\frac{1}{m_r}\int_0^tds\;C_{FF}(t-s)\\
            &=\frac{1}{3\pi^2}\frac{\hbar\,e^2}{\beta^{3}\hbar^3}\operatorname{Im}\left\{\psi^{(2)}(1+i\frac{t}{\beta\hbar})\right\}\,,
\end{align}
where $\psi^{(n)}(z)$ is the $n$th derivative of the digamma
function $\psi(z)$. Again here only the finite temperature
contribution is considered. We see that the mean power is positive
at early times where the stochastic force does positive work to the
charge such that its velocity dispersion or kinetic energy increases
with time. The power reaches the maximum value at the time when
correlation turns negative. Later on, the negative correlation makes
the work done by the force less positive and then the mean power
$\overline{P}_s(t)$ eventually approaches zero,
\begin{align*}
    \overline{P}_{s}(t)&=\frac{1}{3\pi^2m_r}\frac{\hbar\,e^2}{\beta^{3}\hbar^3}
                            \begin{cases}
                                \dfrac{\pi^4}{15}\dfrac{t}{\beta\hbar}+\mathcal{O}(\dfrac{t}{\beta\hbar})^3\,, &t\ll\beta \hbar\,,\\
                            \\
                                \dfrac{\beta^3\hbar^3}{t^3}+\mathcal{O}(\dfrac{\beta\hbar}{t})^5\,,&t\gg\beta\hbar\,.
                        \end{cases}
\end{align*}
Then the finite temperature part of the velocity dispersion $
\langle\Delta v_{i}^{2}(t)\rangle_{\beta}$ can be obtained by
\begin{align*}
      \langle\Delta v_{i}^{2}(t)\rangle_{\beta}&=\frac{2}{m_r}\int_{0}^{t}ds\;\overline{P}_{s}^{i}(s)\\
                        &=\frac{2}{3\pi^2m^2_r}\frac{\hbar\,e^2}{\beta^{2}\hbar^2}\left[\frac{\pi^2}{6}-\operatorname{Re}\psi^{(1)}(1+i\frac{t}{\beta\hbar})\right]\\
                    &=\frac{2}{3\pi^2m^2_r}\frac{\hbar\,e^2}{\beta^{2}\hbar^2}
                        \begin{cases}
                                \dfrac{\pi^4}{30}\dfrac{t^2}{\beta^2\hbar^2}+\mathcal{O}(\dfrac{t}{\beta\hbar})^4\,, &t\ll\beta\hbar\,,\\\\
                            \dfrac{\pi^2}{6}+\mathcal{O}(\dfrac{\beta\hbar}{t})^2\,,&t\gg\beta\hbar\,.
                        \end{cases}
\end{align*}
\begin{figure}
    \centering
        \scalebox{0.75}{\includegraphics{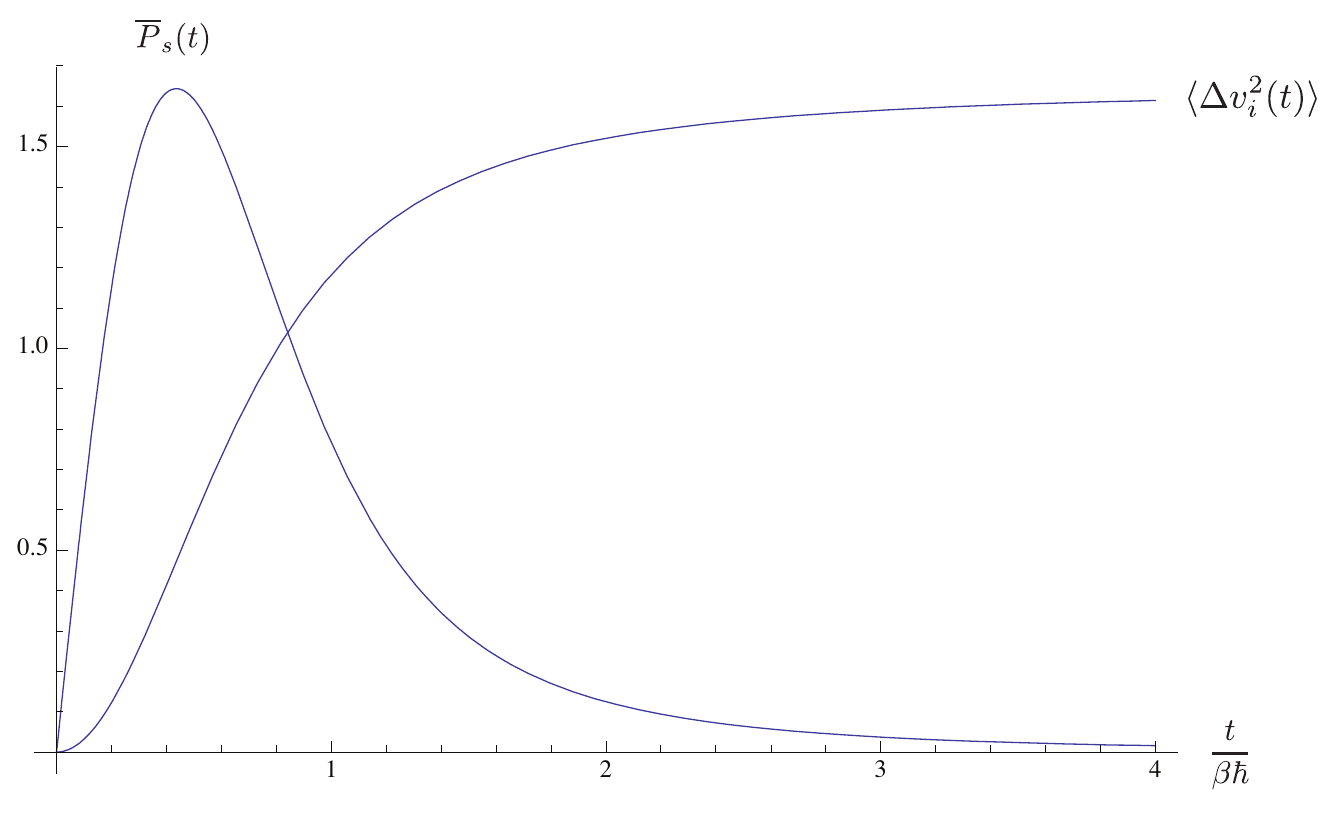}}
        \caption{The time evolution of $\overline{P}_{s} (t) $ and $\langle\Delta v_{i}^{2}(t)\rangle_{\beta}$ is plotted where the time $t$ is in units of $\frac{1}{\beta \hbar}$. These quantities are normalized with $\frac{1}{3\pi^2m_r}\frac{e^2}{\beta^3\hbar^2}$ and $\frac{2}{3\pi^2m_r^2}\frac{e^2}{\beta^2\hbar}$ respectively}\label{Fi:cors}
\end{figure}

In Fig.~\ref{Fi:cors} we show the time evolution of
$\overline{P}_{s}(t)$ and $\langle\Delta
v_{i}^{2}(t)\rangle_{\beta}$ respectively. It is seen that the
averaged power $\overline{P}_{s}(t)$ given by stochastic forces on
the charged particle is positive for all times. This implies that
the thermal bath keeps pumping energy to the particle during its
evolution and increases the velocity dispersion. The presence of the
negative correlation slows down the rate of energy transfer, and
finally leads to the vanishing $\overline{P}_{s}$ at asymptotic
times where velocity fluctuations are saturated. It implies that the
integral of $C_{FF}(\tau)$ over the whole time domain vanishes, i.e.
\begin{equation}\label{inteCFF}
    \int_{0}^{\infty}d\tau\;C_{FF}(\tau)=0=\int_{-\infty}^{\infty}d\tau\;C_{FF}(\tau)\,.
\end{equation}
Thus, the contribution from the positive correlation for the small
$\tau $ regime exactly cancels that from the negative correlation
for the larger $\tau$ regime. It leads to the saturation of the
velocity dispersion even though the self force is insignificant in
this supraohmic case.

To understand Eq.~\eqref{inteCFF} and its consequence from a
different aspect, we may rewrite the correlation function in terms
of its Fourier transform
\begin{align*}
     C_{FF} (\tau)=-\int_0^{\infty}\frac{dk}{2\pi}\;k^{2}\rho^{ii}(k)\left(\frac{1}{e^{\beta\hbar k}-1}\right)\,e^{-ik\tau}+\text{c.c.}\,.
\end{align*}
Then the integration of the correlation function over the whole time domain  becomes
\begin{align}
    \int_{0}^{\infty}d\tau\;C_{FF}(\tau)&=-\int_0^{\infty}dk\;k^{2}\rho^{ii}(k)\left(\frac{1}{e^{\beta\hbar k}-1}\right)\delta(k)\notag\\
                    &=-\frac{1}{2\beta\hbar}\lim_{k\to0}k\rho^{ii}(k)\label{E:coorint}
\end{align}
If $k\rho^{ii}(k)$ behaves like $k^{l-1}$ with $l-1>0$, then
Eq.~\eqref{inteCFF}  holds.  Thus, whether the integral of the
correlation function over the whole time domain vanishes or not
relies on the behavior of $k\rho^{ii}(k)$ in the zero momentum
limit, which in turn depends on the spacetime dimension of the
system, the dispersion relation of the environment field, and the
coupling between the particle and environment.

In electrodynamics, the dynamics of a charged particle is governed
by a local, gauge invariant interaction with the electromagnetic
fields. Under the prescription of minimal coupling, the transverse
component of the vector potential couples to the charged current
density, which is proportional to the time derivative of particle's
position. This derivative coupling gives rise to the fact that the
electromagnetic self-force should be given by a third-order time
derivative of the position, and it is called supraohmic
dynamics~\cite{CAL,HPZ}. From Eq.\eqref{speden}, we see that $l=3$
and hence Eq.~\eqref{inteCFF} holds. In this supraohmic case, the
saturation of velocity dispersion can be achieved due to negative
force-force correlations even when the backreaction effect of the
self-force is found negligible in the evolution.

This is in striking contrast with the Brownian motion under an ohmic environment, characterized by the dissipative backreaction of the first-order time derivative of particle's
position. It involves the coordinate coupling of the particle with the environment~\cite{CAL,HPZ}, and then $l=1$. The positive force-force correlation~\cite{PA} has been shown to drive the growth of the velocity dispersion linearly in time within a time scale shorter than the relaxation time. It is also found~\cite{GZ} that the integration on the force-force correlation function for the whole time regime then has a nonzero value with $l=1$, so the energy transferred from the environment to the particle during the whole evolution of
particle's motion never slows down to halt.  Then the dissipative backreaction must be taken into account to counterbalance the effect from the force fluctuations in order to finally stabilize the value of the velocity dispersion. Moreover, from the aspect of energy
balancing between dissipative and fluctuation backreations, the stronger dissipation is expected to occur in the subohmic case with $l<1$, because the energy transfer rate increases even faster at late times~\cite{HPZ}. Thus, the mechanisms to stabilize the velocity dispersion of a particle in a fluctuating subohmic, ohmic, or supraohmic environment are rather different.

\section{summary and concluding remarks}
In this paper, the evolution of velocity fluctuations of the
Brownian particle in a supraohmic environment is studied by
considering the motion of the charged particle  coupled to
electromagnetic fluctuations at finite temperature. The Langevin
equation of the particle incorporates the effects of fluctuation and
dissipation backreaction in a self-consistent manner. In particular,
the backreaction in a form of the electromagnetic self-force on the
charge is a third-order time derivative of the position in this
supraohmic environment. On the other hand, the thermal fluctuations
of the electromagnetic fields are manifested as stochastic noise.
Its correlation tends to be positive at shorter time difference
$\tau$, and then turns negative at $\tau\sim\mathcal{O}(\beta)$. We
show that the integration of the force-force correlation function
over the whole time regime vanishes. Throughout the evolution, the
self-force can be argued to be insignificant for the charge having
no significant motion. Then, the positive correlation contributes to
the growth of the velocity dispersion initially, its growth slows
down when correlation becomes negative, and finally halts where the
velocity dispersion reaches a constant at asymptotical times. It is
a rather different saturation mechanism for the velocity dispersion
of the Brownian particle in an ohmic environment.

Additionally, the saturated value of the kinetic energy of the
charge due to the thermal electromagnetic fluctuations at
temperature $T=\beta^{-1}$ can be estimated by
\begin{equation}
   \frac{1}{2} m_r  \langle\Delta v_i^2(\infty)\rangle_{\beta}
   \sim\alpha\,\frac{k_BT}{m_{r}c^2}\,k_B T \sim 10^{-5} \left( \frac{ k_B
   T }{k eV}\right)^2 keV\,,
\end{equation}
with the fine structure constant $\alpha=e^2/\hbar c$. The $\alpha$
dependence implies that the saturated value of the velocity
dispersion in this supraohmic case relies on the coupling between
the system and the environment. The issue on  how the velocity
dispersion of the Brownian particle is stabilized in the subohmic
environment deserves further investigation.

Thus, it comes to no surprise that the above saturated value of the
velocity dispersion roughly becomes the value given by the
equipartition theorem for a thermodynamic system as long as $k_B T
\sim m_r c^2 /\alpha \sim  \hbar c \tau_{e}^{-1}$  because in that
limit, the temperature is the highest energy scale in the system.
However, in order to correctly describe the system in such a limit,
a relativistic/field-theoretic formalism is needed  for
appropriately describing the dynamics of the charged particle. This
is beyond the scope of the current investigation.

\begin{acknowledgments}
We would like to thank Larry H. Ford and C.-H. Wu for stimulating discussions. This work was supported in part by the National Science Council, R. O. C. under grant NSC95-2112-M-259-011-MY2, and the National Center for Theoretical Sciences, Taiwan.
\end{acknowledgments}

\end{document}